# The Growing Liberality Observed in Primary Animal and Plant Cultures is Common to the Social Amoeba

Norichika Ogata
Nihon BioData Corporation, Kawasaki, Japan
norichik@nbiodata.com

*Abstract*—**Tissue culture environment liberates cells from ordinary laws of multi-cellular organisms. This liberation enables cells several behaviors, such as proliferation, dedifferentiation, acquisition of pluripotency, immortalization, and reprogramming. Recently, the quantitative value of cellular dedifferentiation and differentiation was defined as "liberality", which is measurable as Shannon entropy of numerical transcriptome data and Lempel-Zip complexity of nucleotide sequence transcriptome data. The increasing liberality induced by the culture environment had first been observed in animal cells and had reconfirmed in plant cells. The phenomena may be common across the kingdom, also in a social amoeba. We measured the liberality of the social amoeba which disaggregated from multicellular aggregates and transferred into a liquid medium.**

*Keywords—Genomics, Transcriptome, Social Amoeba, Dictyostelium, Dedifferentiation, Liberality, Primary Culture, Lempel-Ziv Complexity.*

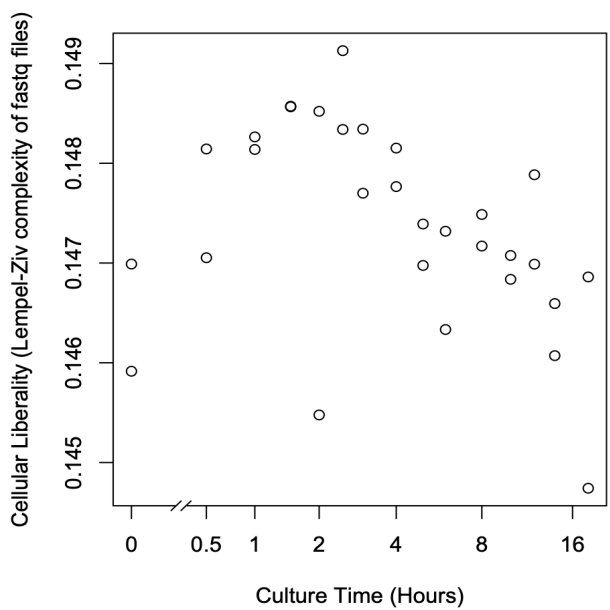

**Fig 1. Scatter plot of culture time vs cellular liberality.**

The social amoeba cells disaggregated of multicellular aggregates and transferred the cells into a HL5 liquid medium. The Lempel-Ziv complexity of transcriptome were measured in each culture time.

## I. INTRODUCTION

Tissue culture is performed to maintain isolated portions of multicellular organisms in an artificial milieu that is outside the individual organism and for considerable periods of time [1]. It is known over a century that cells derived from cultured explants are, in general, different from cells of the corresponding tissue in a living organism [2, 3]. In these tissue cultures, cells are liberated from stimulations and prohibition which is ordinary in multi-cellular organisms [4]. This liberation is essential for growth, dedifferentiation, acquisition of pluripotency, immortalization and reprogramming. The proliferations of cultured cells were also considered to be a result of dedifferentiation [2].

In this decade, several studies [5-8] following our research [9] repeatedly measured the degree of cellular dedifferentiation and differentiation as a Shannon entropy of numerical transcriptome data. The Shannon entropy is a kind of alpha diversity in the ecology [10], and the transcriptome Shannon entropy is simply transcriptome diversity [11]. It is not incorrect to call it the (alpha) diversity of the transcriptome, but that would leave its biological significance undefined, as would each principal component that came up in the principal component analysis. We can quantitatively assess, judge, and define that dedifferentiation is an increase in the Shannon entropy of the transcriptome, it is more accurate to position the value not as a mere bioinformatics measure; however, as a number with obvious biological and bioengineering significance, such as viable cell rate, cell density, specific growth rate, or pcd (pg/cell/day) [12]. Here we call the quantitative value of cellular dedifferentiation and differentiation "liberality," since a previous study explained the changes were happening to cultured cells as "libère" [13]. The study of cellular liberality has entered a phase in which evidence is being gathered to strengthen the theory; the degree of cellular dedifferentiation and differentiation is measurable.

We previously observed increasing liberality in animal and plant primary cultures [5, 14]. In this study, we measured the liberality of the social amoeba, *Dictyostelium discoideum*, having a transition from single-celled amoebae to a multicellular organism as a natural part of its life cycle [15]. *Dictyostelium* dedifferentiation was initiated by disaggregation of multicellular aggregates and transfer of the cells into a HL5 liquid medium in a study [16]. We could not reproduce the mapping used in the study since the referential genome was not clear in the manuscript; There is *"Reads were mapped to the Dictyostelium genome (version obtained from Gareth Bloomfield, masking the duplication on chromosome 2) using Tophat v2.0.9."*. Therefore, we measured the liberality from the transcriptome data without the referential genome and the reads mapping process. A recent study enabled measuring the liberality as Lempel-Ziv complexity of fastq files [17, 18].

## II. Materials and Methods

Transcriptome data set was obtained from DDBJ SRA (SRA1039371) [16]; 30 fastq data ware used (SRR11039842-SRR11039872). In the entry, time course total RNA sampling during cultivation of the amoeba (0, 0.5, 1, 1.5, 2, 2.5, 3, 4, 5, 6, 8, 10, 12, 14 and 18 hours). Each sample has two biological replicates. We downloaded sra compressed fastq data. The file size of decompressed raw textual fastq files and bz2 compressed files were measured. We measured Lempel-Ziv complexity of the fastq files as file size rate between bz2 compressed fastq data and raw textual fastq data. Then we compared the culture time and the Lampel-Ziv complexity.

## III. Results and Discussion

*Dictyostelium* dedifferentiation is initiated by disaggregation of multicellular aggregates and transfer of the cells into a HL5 liquid media. The liberality immediately increased and reached the peak in 3 hours. This result suggested that dedifferentiation of cells in primary explant culture is a common phenomenon for diverse multi-cellular organisms including social amoeba. After 4 hours of the amoeba disaggregation, the liberality started decreasing. The liberality of the amoeba 18 hours after disaggregation were equal to 0 hours after disaggregation. This liberality decreasing following the liberality increase was observed in the previous study of plants, that was thought to be re-differentiation to construct a new plant. The dedifferentiated amoeba would immediately start re-differentiation to achieve a slightly better position in the new environment. Although, in the *Dictyostelium* paper the amoeba had been thought to dedifferentiate for 18 hours, our observations of the liberality suggest that amoeba dedifferentiation completed in 4 hours. Indeed, the *Dictyostelim* paper indicated the importantest change of amoeba at 2~4 hours after disaggregation; two transcription factors, *bzpS* and *mybD* showed clear peak of expression (Fig. 3a [16]), the PC1 in a primary component analysis of transcriptome does not change after 4 hours after disaggregation (Fig. 3b [16]), and cell motility showed peak at 3 hours after disaggregation (Fig. 5 [16]).

Even plants complete dedifferentiation within 6 hours [14, 19], so there is no reason why the amoeba dedifferentiation should be slower than that. We recommend that studies attempting to examine dedifferentiation incorporate measurements of cellular liberality to accurately monitor the state of cellular dedifferentiation [20]. As was evident in the present study, measurement processes that include mapping of reads to a reference genome are less reproducible, making the measurement of liberality based on LZ complexity of nucleotide sequence files [18] more useful.

Appendix. Codes obtaining data used in this study.

# GSM4300090: Media 0h rep1; Dictyostelium discoideum; RNA-Seq
wget ftp://ftp.ddbj.nig.ac.jp/ddbj_database/dra/sralite/ByExp/litesra/SRX/SRX769/SRX7691764/SRR11039842/SRR11039842.sra ./

# GSM4300091: Media 0.5h rep1; Dictyostelium discoideum; RNA-Seq
wget ftp://ftp.ddbj.nig.ac.jp/ddbj_database/dra/sralite/ByExp/litesra/SRX/SRX769/SRX7691765/SRR11039843/SRR11039843.sra ./

# GSM4300092: Media 1h rep1; Dictyostelium discoideum; RNA-Seq
wget ftp://ftp.ddbj.nig.ac.jp/ddbj_database/dra/sralite/ByExp/litesra/SRX/SRX769/SRX7691766/SRR11039844/SRR11039844.sra ./

# GSM4300093: Media 1.5h rep1; Dictyostelium discoideum; RNA-Seq
wget ftp://ftp.ddbj.nig.ac.jp/ddbj_database/dra/sralite/ByExp/litesra/SRX/SRX769/SRX7691767/SRR11039846/SRR11039846.sra ./

# GSM4300094: Media 2h rep1; Dictyostelium discoideum; RNA-Seq
wget ftp://ftp.ddbj.nig.ac.jp/ddbj_database/dra/sralite/ByExp/litesra/SRX/SRX769/SRX7691768/SRR11039847/SRR11039847.sra ./

# GSM4300095: Media 2.5h rep1; Dictyostelium discoideum; RNA-Seq
wget ftp://ftp.ddbj.nig.ac.jp/ddbj_database/dra/sralite/ByExp/litesra/SRX/SRX769/SRX7691770/SRR11039848/SRR11039848.sra ./

# GSM4300096: Media 3h rep1; Dictyostelium discoideum; RNA-Seq
wget ftp://ftp.ddbj.nig.ac.jp/ddbj_database/dra/sralite/ByExp/litesra/SRX/SRX769/SRX7691771/SRR11039849/SRR11039849.sra ./

# GSM4300097: Media 4h rep1; Dictyostelium discoideum; RNA-Seq
wget ftp://ftp.ddbj.nig.ac.jp/ddbj_database/dra/sralite/ByExp/litesra/SRX/SRX769/SRX7691772/SRR11039850/SRR11039850.sra ./

# GSM4300098: Media 5h rep1; Dictyostelium discoideum; RNA-Seq
wget ftp://ftp.ddbj.nig.ac.jp/ddbj_database/dra/sralite/ByExp/litesra/SRX/SRX769/SRX7691773/SRR11039851/SRR11039851.sra ./

# GSM4300099: Media 6h rep1; Dictyostelium discoideum; RNA-Seq
wget ftp://ftp.ddbj.nig.ac.jp/ddbj_database/dra/sralite/ByExp/litesra/SRX/SRX769/SRX7691774/SRR11039852/SRR11039852.sra ./

# GSM4300100: Media 8h rep1; Dictyostelium discoideum; RNA-Seq
wget ftp://ftp.ddbj.nig.ac.jp/ddbj_database/dra/sralite/ByExp/litesra/SRX/SRX769/SRX7691775/SRR11039853/SRR11039853.sra ./

# GSM4300101: Media 10h rep1; Dictyostelium discoideum; RNA-Seq
wget ftp://ftp.ddbj.nig.ac.jp/ddbj_database/dra/sralite/ByExp/litesra/SRX/SRX769/SRX7691776/SRR11039854/SRR11039854.sra ./

# GSM4300102: Media 12h rep1; Dictyostelium discoideum; RNA-Seq
wget ftp://ftp.ddbj.nig.ac.jp/ddbj_database/dra/sralite/ByExp/litesra/SRX/SRX769/SRX7691777/SRR11039855/SRR11039855.sra ./

# GSM4300103: Media 14h rep1; Dictyostelium discoideum; RNA-Seq
wget ftp://ftp.ddbj.nig.ac.jp/ddbj_database/dra/sralite/ByExp/litesra/SRX/SRX769/SRX7691778/SRR11039856/SRR11039856.sra ./

# GSM4300104: Media 18h rep1; Dictyostelium discoideum; RNA-Seq
wget ftp://ftp.ddbj.nig.ac.jp/ddbj_database/dra/sralite/ByExp/litesra/SRX/SRX769/SRX7691779/SRR11039857/SRR11039857.sra ./

# GSM4300105: Media 0h rep2; Dictyostelium discoideum; RNA-Seq
wget ftp://ftp.ddbj.nig.ac.jp/ddbj_database/dra/sralite/ByExp/litesra/SRX/SRX769/SRX7691780/SRR11039858/SRR11039858.sra ./

# GSM4300106: Media 0.5h rep2; Dictyostelium discoideum; RNA-Seq
wget ftp://ftp.ddbj.nig.ac.jp/ddbj_database/dra/sralite/ByExp/litesra/SRX/SRX769/SRX7691781/SRR11039859/SRR11039859.sra ./

# GSM4300107: Media 1h rep2; Dictyostelium discoideum; RNA-Seq
wget ftp://ftp.ddbj.nig.ac.jp/ddbj_database/dra/sralite/ByExp/litesra/SRX/SRX769/SRX7691782/SRR11039860/SRR11039860.sra ./

# GSM4300108: Media 1.5h rep2; Dictyostelium discoideum; RNA-Seq
wget ftp://ftp.ddbj.nig.ac.jp/ddbj_database/dra/sralite/ByExp/litesra/SRX/SRX769/SRX7691783/SRR11039861/SRR11039861.sra ./

# GSM4300109: Media 2h rep2; Dictyostelium discoideum; RNA-Seq
wget ftp://ftp.ddbj.nig.ac.jp/ddbj_database/dra/sralite/ByExp/litesra/SRX/SRX769/SRX7691784/SRR11039862/SRR11039862.sra ./

# GSM4300110: Media 2.5h rep2; Dictyostelium discoideum; RNA-Seq
wget ftp://ftp.ddbj.nig.ac.jp/ddbj_database/dra/sralite/ByExp/litesra/SRX/SRX769/SRX7691785/SRR11039863/SRR11039863.sra ./

# GSM4300111: Media 3h rep2; Dictyostelium discoideum; RNA-Seq
wget ftp://ftp.ddbj.nig.ac.jp/ddbj_database/dra/sralite/ByExp/litesra/SRX/SRX769/SRX7691786/SRR11039864/SRR11039864.sra ./

# GSM4300112: Media 4h rep2; Dictyostelium discoideum; RNA-Seq
wget ftp://ftp.ddbj.nig.ac.jp/ddbj_database/dra/sralite/ByExp/litesra/SRX/SRX769/SRX7691787/SRR11039865/SRR11039865.sra ./

# GSM4300113: Media 5h rep2; Dictyostelium discoideum; RNA-Seq
wget ftp://ftp.ddbj.nig.ac.jp/ddbj_database/dra/sralite/ByExp/litesra/SRX/SRX769/SRX7691788/SRR11039866/SRR11039866.sra ./

# GSM4300114: Media 6h rep2; Dictyostelium discoideum; RNA-Seq
wget ftp://ftp.ddbj.nig.ac.jp/ddbj_database/dra/sralite/ByExp/litesra/SRX/SRX769/SRX7691801/SRR11039867/SRR11039867.sra ./

# GSM4300115: Media 8h rep2; Dictyostelium discoideum; RNA-Seq
wget ftp://ftp.ddbj.nig.ac.jp/ddbj_database/dra/sralite/ByExp/litesra/SRX/SRX769/SRX7691802/SRR11039868/SRR11039868.sra ./

# GSM4300116: Media 10h rep2; Dictyostelium discoideum; RNA-Seq
wget ftp://ftp.ddbj.nig.ac.jp/ddbj_database/dra/sralite/ByExp/litesra/SRX/SRX769/SRX7691803/SRR11039869/SRR11039869.sra ./

# GSM4300117: Media 12h rep2; Dictyostelium discoideum; RNA-Seq
wget ftp://ftp.ddbj.nig.ac.jp/ddbj_database/dra/sralite/ByExp/litesra/SRX/SRX769/SRX7691804/SRR11039870/SRR11039870.sra ./

# GSM4300118: Media 14h rep2; Dictyostelium discoideum; RNA-Seq
wget ftp://ftp.ddbj.nig.ac.jp/ddbj_database/dra/sralite/ByExp/litesra/SRX/SRX769/SRX7691805/SRR11039871/SRR11039871.sra ./

# GSM4300119: Media 18h rep2; Dictyostelium discoideum; RNA-Seq
wget ftp://ftp.ddbj.nig.ac.jp/ddbj_database/dra/sralite/ByExp/litesra/SRX/SRX769/SRX7691806/SRR11039872/SRR11039872.sra ./

# GSM4300120: Undifferentiated rep1; Dictyostelium discoideum; RNA-Seq
wget ftp://ftp.ddbj.nig.ac.jp/ddbj_database/dra/sralite/ByExp/litesra/SRX/SRX769/SRX7691807/SRR11039873/SRR11039873.sra ./

# GSM4300121: Undifferentiated rep2; Dictyostelium discoideum; RNA-Seq
wget ftp://ftp.ddbj.nig.ac.jp/ddbj_database/dra/sralite/ByExp/litesra/SRX/SRX769/SRX7691808/SRR11039874/SRR11039874.sra ./

# GSM4300122: Bacteria/Buffer 0h rep1; Dictyostelium discoideum; RNA-Seq
wget ftp://ftp.ddbj.nig.ac.jp/ddbj_database/dra/sralite/ByExp/litesra/SRX/SRX769/SRX7691809/SRR11039875/SRR11039875.sra ./

# GSM4300123: Bacteria 0.5h rep1; Dictyostelium discoideum; RNA-Seq
wget ftp://ftp.ddbj.nig.ac.jp/ddbj_database/dra/sralite/ByExp/litesra/SRX/SRX769/SRX7691810/SRR11039876/SRR11039876.sra ./

# GSM4300124: Bacteria 1h rep1; Dictyostelium discoideum; RNA-Seq
wget ftp://ftp.ddbj.nig.ac.jp/ddbj_database/dra/sralite/ByExp/litesra/SRX/SRX769/SRX7691811/SRR11039877/SRR11039877.sra ./

# GSM4300125: Bacteria 1.5h rep1; Dictyostelium discoideum; RNA-Seq
wget ftp://ftp.ddbj.nig.ac.jp/ddbj_database/dra/sralite/ByExp/litesra/SRX/SRX769/SRX7691812/SRR11039878/SRR11039878.sra ./

# GSM4300126: Bacteria 2h rep1; Dictyostelium discoideum; RNA-Seq
wget ftp://ftp.ddbj.nig.ac.jp/ddbj_database/dra/sralite/ByExp/litesra/SRX/SRX769/SRX7691814/SRR11039879/SRR11039879.sra ./

# GSM4300127: Bacteria 3h rep1; Dictyostelium discoideum; RNA-Seq
wget ftp://ftp.ddbj.nig.ac.jp/ddbj_database/dra/sralite/ByExp/litesra/SRX/SRX769/SRX7691815/SRR11039880/SRR11039880.sra ./

# GSM4300128: Bacteria 4h rep1; Dictyostelium discoideum; RNA-Seq
wget ftp://ftp.ddbj.nig.ac.jp/ddbj_database/dra/sralite/ByExp/litesra/SRX/SRX769/SRX7691816/SRR11039881/SRR11039881.sra ./

# GSM4300129: Bacteria 5h rep1; Dictyostelium discoideum; RNA-Seq
wget ftp://ftp.ddbj.nig.ac.jp/ddbj_database/dra/sralite/ByExp/litesra/SRX/SRX769/SRX7691817/SRR11039882/SRR11039882.sra ./

# GSM4300130: Bacteria 6h rep1; Dictyostelium discoideum; RNA-Seq
wget ftp://ftp.ddbj.nig.ac.jp/ddbj_database/dra/sralite/ByExp/litesra/SRX/SRX769/SRX7691818/SRR11039883/SRR11039883.sra ./

# GSM4300131: Bacteria 8h rep1; Dictyostelium discoideum; RNA-Seq
wget ftp://ftp.ddbj.nig.ac.jp/ddbj_database/dra/sralite/ByExp/litesra/SRX/SRX769/SRX7691819/SRR11039884/SRR11039884.sra ./

# GSM4300132: Bacteria 12h rep1; Dictyostelium discoideum; RNA-Seq
wget ftp://ftp.ddbj.nig.ac.jp/ddbj_database/dra/sralite/ByExp/litesra/SRX/SRX769/SRX7691820/SRR11039885/SRR11039885.sra ./

# GSM4300133: Bacteria 24h rep1; Dictyostelium discoideum; RNA-Seq
wget ftp://ftp.ddbj.nig.ac.jp/ddbj_database/dra/sralite/ByExp/litesra/SRX/SRX769/SRX7691821/SRR11039886/SRR11039886.sra ./

# GSM4300134: Buffer 0.5h rep1; Dictyostelium discoideum; RNA-Seq
wget ftp://ftp.ddbj.nig.ac.jp/ddbj_database/dra/sralite/ByExp/litesra/SRX/SRX769/SRX7691822/SRR11039887/SRR11039887.sra ./

# GSM4300135: Buffer 1h rep1; Dictyostelium discoideum; RNA-Seq
wget ftp://ftp.ddbj.nig.ac.jp/ddbj_database/dra/sralite/ByExp/litesra/SRX/SRX769/SRX7691823/SRR11039888/SRR11039888.sra ./

# GSM4300136: Buffer 2h rep1; Dictyostelium discoideum; RNA-Seq
wget ftp://ftp.ddbj.nig.ac.jp/ddbj_database/dra/sralite/ByExp/litesra/SRX/SRX769/SRX7691824/SRR11039889/SRR11039889.sra ./

# GSM4300137: Buffer 3h rep1; Dictyostelium discoideum; RNA-Seq
wget ftp://ftp.ddbj.nig.ac.jp/ddbj_database/dra/sralite/ByExp/litesra/SRX/SRX769/SRX7691825/SRR11039890/SRR11039890.sra ./

# GSM4300138: Buffer 4h rep1; Dictyostelium discoideum; RNA-Seq
wget ftp://ftp.ddbj.nig.ac.jp/ddbj_database/dra/sralite/ByExp/litesra/SRX/SRX769/SRX7691826/SRR11039891/SRR11039891.sra ./

# GSM4300139: Buffer 6h rep1; Dictyostelium discoideum; RNA-Seq
wget ftp://ftp.ddbj.nig.ac.jp/ddbj_database/dra/sralite/ByExp/litesra/SRX/SRX769/SRX7691827/SRR11039892/SRR11039892.sra ./

# GSM4300140: Bacteria/Buffer 0h rep2; Dictyostelium discoideum; RNA-Seq
wget ftp://ftp.ddbj.nig.ac.jp/ddbj_database/dra/sralite/ByExp/litesra/SRX/SRX769/SRX7691828/SRR11039893/SRR11039893.sra ./

# GSM4300141: Bacteria 0.5h rep2; Dictyostelium discoideum; RNA-Seq
wget ftp://ftp.ddbj.nig.ac.jp/ddbj_database/dra/sralite/ByExp/litesra/SRX/SRX769/SRX7691829/SRR11039894/SRR11039894.sra ./

# GSM4300142: Bacteria 1h rep2; Dictyostelium discoideum; RNA-Seq
wget ftp://ftp.ddbj.nig.ac.jp/ddbj_database/dra/sralite/ByExp/litesra/SRX/SRX769/SRX7691830/SRR11039895/SRR11039895.sra ./

# GSM4300143: Bacteria 1.5h rep2; Dictyostelium discoideum; RNA-Seq
wget ftp://ftp.ddbj.nig.ac.jp/ddbj_database/dra/sralite/ByExp/litesra/SRX/SRX769/SRX7691831/SRR11039896/SRR11039896.sra ./

# GSM4300144: Bacteria 2h rep2; Dictyostelium discoideum; RNA-Seq
wget ftp://ftp.ddbj.nig.ac.jp/ddbj_database/dra/sralite/ByExp/litesra/SRX/SRX769/SRX7691832/SRR11039897/SRR11039897.sra ./

# GSM4300145: Bacteria 3h rep2; Dictyostelium discoideum; RNA-Seq
wget ftp://ftp.ddbj.nig.ac.jp/ddbj_database/dra/sralite/ByExp/litesra/SRX/SRX769/SRX7691833/SRR11039898/SRR11039898.sra ./

# GSM4300146: Bacteria 4h rep2; Dictyostelium discoideum; RNA-Seq
wget ftp://ftp.ddbj.nig.ac.jp/ddbj_database/dra/sralite/ByExp/litesra/SRX/SRX769/SRX7691834/SRR11039899/SRR11039899.sra ./

# GSM4300147: Bacteria 5h rep2; Dictyostelium discoideum; RNA-Seq
wget ftp://ftp.ddbj.nig.ac.jp/ddbj_database/dra/sralite/ByExp/litesra/SRX/SRX769/SRX7691835/SRR11039900/SRR11039900.sra ./

# GSM4300148: Bacteria 6h rep2; Dictyostelium discoideum; RNA-Seq
wget ftp://ftp.ddbj.nig.ac.jp/ddbj_database/dra/sralite/ByExp/litesra/SRX/SRX769/SRX7691836/SRR11039901/SRR11039901.sra ./

# GSM4300149: Bacteria 8h rep2; Dictyostelium discoideum; RNA-Seq
wget ftp://ftp.ddbj.nig.ac.jp/ddbj_database/dra/sralite/ByExp/litesra/SRX/SRX769/SRX7691837/SRR11039902/SRR11039902.sra ./

# GSM4300150: Bacteria 12h rep2; Dictyostelium discoideum; RNA-Seq
wget ftp://ftp.ddbj.nig.ac.jp/ddbj_database/dra/sralite/ByExp/litesra/SRX/SRX769/SRX7691838/SRR11039903/SRR11039903.sra ./

# GSM4300151: Bacteria 24h rep2; Dictyostelium discoideum; RNA-Seq
wget ftp://ftp.ddbj.nig.ac.jp/ddbj_database/dra/sralite/ByExp/litesra/SRX/SRX769/SRX7691839/SRR11039904/SRR11039904.sra ./

# GSM4300152: Buffer 0.5h rep2; Dictyostelium discoideum; RNA-Seq
wget ftp://ftp.ddbj.nig.ac.jp/ddbj_database/dra/sralite/ByExp/litesra/SRX/SRX769/SRX7691840/SRR11039906/SRR11039906.sra ./

# GSM4300153: Buffer 1h rep2; Dictyostelium discoideum; RNA-Seq
wget ftp://ftp.ddbj.nig.ac.jp/ddbj_database/dra/sralite/ByExp/litesra/SRX/SRX769/SRX7691841/SRR11039907/SRR11039907.sra ./

# GSM4300154: Buffer 2h rep2; Dictyostelium discoideum; RNA-Seq
wget ftp://ftp.ddbj.nig.ac.jp/ddbj_database/dra/sralite/ByExp/litesra/SRX/SRX769/SRX7691842/SRR11039908/SRR11039908.sra ./

# GSM4300155: Buffer 3h rep2; Dictyostelium discoideum; RNA-Seq
wget ftp://ftp.ddbj.nig.ac.jp/ddbj_database/dra/sralite/ByExp/litesra/SRX/SRX769/SRX7691843/SRR11039909/SRR11039909.sra ./

# GSM4300156: Buffer 4h rep2; Dictyostelium discoideum; RNA-Seq
wget ftp://ftp.ddbj.nig.ac.jp/ddbj_database/dra/sralite/ByExp/litesra/SRX/SRX769/SRX7691844/SRR11039910/SRR11039910.sra ./

# GSM4300157: Buffer 6h rep2; Dictyostelium discoideum; RNA-Seq
wget ftp://ftp.ddbj.nig.ac.jp/ddbj_database/dra/sralite/ByExp/litesra/SRX/SRX769/SRX7691846/SRR11039911/SRR11039911.sra ./

# GSM4300158: Development 0h rep1; Dictyostelium discoideum; RNA-Seq
wget ftp://ftp.ddbj.nig.ac.jp/ddbj_database/dra/sralite/ByExp/litesra/SRX/SRX769/SRX7691847/SRR11039912/SRR11039912.sra ./

# GSM4300159: Development 2h rep1; Dictyostelium discoideum; RNA-Seq
wget ftp://ftp.ddbj.nig.ac.jp/ddbj_database/dra/sralite/ByExp/litesra/SRX/SRX769/SRX7691848/SRR11039913/SRR11039913.sra ./

# GSM4300160: Development 4h rep1; Dictyostelium discoideum; RNA-Seq
wget ftp://ftp.ddbj.nig.ac.jp/ddbj_database/dra/sralite/ByExp/litesra/SRX/SRX769/SRX7691849/SRR11039914/SRR11039914.sra ./

# GSM4300161: Development 6h rep1; Dictyostelium discoideum; RNA-Seq
wget ftp://ftp.ddbj.nig.ac.jp/ddbj_database/dra/sralite/ByExp/litesra/SRX/SRX769/SRX7691850/SRR11039915/SRR11039915.sra ./

# GSM4300162: Development 8h rep1; Dictyostelium discoideum; RNA-Seq
wget ftp://ftp.ddbj.nig.ac.jp/ddbj_database/dra/sralite/ByExp/litesra/SRX/SRX769/SRX7691789/SRR11039916/SRR11039916.sra ./

# GSM4300163: Development 10h rep1; Dictyostelium discoideum; RNA-Seq
wget ftp://ftp.ddbj.nig.ac.jp/ddbj_database/dra/sralite/ByExp/litesra/SRX/SRX769/SRX7691790/SRR11039917/SRR11039917.sra ./

# GSM4300164: Development 12h rep1; Dictyostelium discoideum; RNA-Seq
wget ftp://ftp.ddbj.nig.ac.jp/ddbj_database/dra/sralite/ByExp/litesra/SRX/SRX769/SRX7691791/SRR11039918/SRR11039918.sra ./

# GSM4300165: Development 14h rep1; Dictyostelium discoideum; RNA-Seq
wget ftp://ftp.ddbj.nig.ac.jp/ddbj_database/dra/sralite/ByExp/litesra/SRX/SRX769/SRX7691792/SRR11039919/SRR11039919.sra ./

# GSM4300166: Development 0h rep2; Dictyostelium discoideum; RNA-Seq
wget ftp://ftp.ddbj.nig.ac.jp/ddbj_database/dra/sralite/ByExp/litesra/SRX/SRX769/SRX7691793/SRR11039920/SRR11039920.sra ./

# GSM4300167: Development 2h rep2; Dictyostelium discoideum; RNA-Seq
wget ftp://ftp.ddbj.nig.ac.jp/ddbj_database/dra/sralite/ByExp/litesra/SRX/SRX769/SRX7691794/SRR11039921/SRR11039921.sra ./

# GSM4300168: Development 4h rep2; Dictyostelium discoideum; RNA-Seq
wget ftp://ftp.ddbj.nig.ac.jp/ddbj_database/dra/sralite/ByExp/litesra/SRX/SRX769/SRX7691795/SRR11039922/SRR11039922.sra ./

# GSM4300169: Development 6h rep2; Dictyostelium discoideum; RNA-Seq
wget ftp://ftp.ddbj.nig.ac.jp/ddbj_database/dra/sralite/ByExp/litesra/SRX/SRX769/SRX7691796/SRR11039923/SRR11039923.sra ./

# GSM4300170: Development 8h rep2; Dictyostelium discoideum; RNA-Seq
wget ftp://ftp.ddbj.nig.ac.jp/ddbj_database/dra/sralite/ByExp/litesra/SRX/SRX769/SRX7691797/SRR11039924/SRR11039924.sra ./

# GSM4300171: Development 10h rep2; Dictyostelium discoideum; RNA-Seq
wget ftp://ftp.ddbj.nig.ac.jp/ddbj_database/dra/sralite/ByExp/litesra/SRX/SRX769/SRX7691798/SRR11039925/SRR11039925.sra ./

# GSM4300172: Development 12h rep2; Dictyostelium discoideum; RNA-Seq
wget ftp://ftp.ddbj.nig.ac.jp/ddbj_database/dra/sralite/ByExp/litesra/SRX/SRX769/SRX7691799/SRR11039926/SRR11039926.sra ./

# GSM4300173: Development 14h rep2; Dictyostelium discoideum; RNA-Seq
wget ftp://ftp.ddbj.nig.ac.jp/ddbj_database/dra/sralite/ByExp/litesra/SRX/SRX769/SRX7691800/SRR11039927/SRR11039927.sra ./